\documentclass[letterpaper,10pt,conference]{ieeeconf}  %
\IEEEoverridecommandlockouts                         
\usepackage{cite}
\usepackage{amsmath,amssymb,amsfonts}
\usepackage{algorithmic}
\usepackage{graphicx}
\usepackage{textcomp}
\usepackage{xcolor}
\usepackage{soul,dsfont} 
\usepackage{tikz} 
\usepackage{color}
\usepackage{algorithm} 
\usepackage{array}
\usepackage{eqparbox}
\usepackage{url}
\usepackage{relsize}
\usepackage{stfloats}
\usepackage{graphics}

\def\BibTeX{{\rm B\kern-.05em{\sc i\kern-.025em b}\kern-.08em
T\kern-.1667em\lower.7ex\hbox{E}\kern-.125emX}}

\title{\LARGE \bf
On Complexity of Stability Analysis in Higher-order Ecological Networks through Tensor Decompositions
}

\author{Anqi Dong$^{1}$ and Can Chen$^{2}$
\thanks{$^{1}$Anqi Dong is with the Department of Mechanical and Aerospace Engineering, University of California, Irvine, Irvine, CA 92617, USA. 
{\tt\small anqid2@uci.edu}}%
\thanks{$^{2}$Can Chen is with the School of Data Science and Society and the Department of Mathematics, University of North Carolina at Chapel Hill, Chapel Hill, NC 27599, USA.
{\tt\small canc@unc.edu}}%
}

\begin{document}

\maketitle
\thispagestyle{empty}
\pagestyle{empty}

\begin{abstract}
Complex ecological networks are often characterized by intricate interactions that extend beyond pairwise relationships. Understanding the stability of higher-order ecological networks is salient for species coexistence, biodiversity, and community persistence. In this article, we present complexity analyses for determining the linear stability of higher-order ecological networks through tensor decompositions. We are interested in the higher-order generalized Lotka-Volterra model, which captures high-order interactions using tensors of varying orders. To efficiently compute Jacobian matrices and thus determine stability in large ecological networks, we exploit various tensor decompositions, including higher-order singular value decomposition, Canonical Polyadic decomposition, and tensor train decomposition, accompanied by in-depth computational and memory complexity analyses. We demonstrate the effectiveness of our framework with numerical examples. 
\end{abstract}
\vspace{0.1in}
\begin{keywords}
Linear stability, Jacobian matrices, ecological networks, higher-order interactions, tensor decomposition. 
\end{keywords}

\section{Introduction}
Modeling complex systems and predicting their long-term behavior have drawn significant attention across 
diverse fields, including mathematics, physics, biology, and social science \cite{skardal2020higher,dong2023negative}. Network representations, which assign individual components to nodes and their connections to edges, are widely used to characterize complex systems. They place a specific emphasis on understanding the influence of interactions, quantified by the edge weights, as in weighted Laplacian or adjacency matrices. Networks have been proven to be powerful tools in system modeling, with numerous works on related system-theoretic properties such as stability, controllability, and observability.

\begin{figure}[t]
    \centering
    \includegraphics[width=0.28\textwidth]{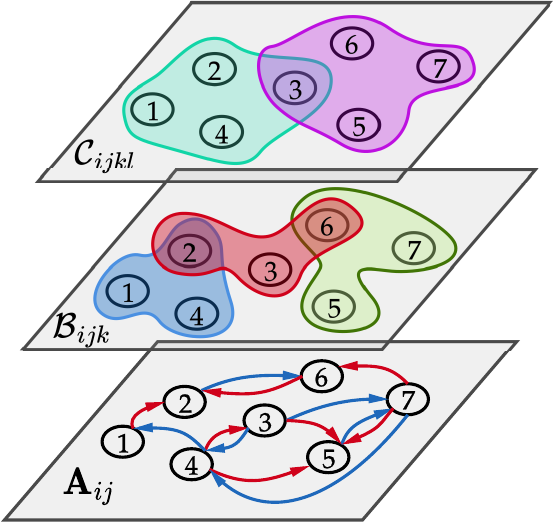}
    \caption{High-order network representation of an ecological system with seven species, where interactions can occur between pairs or groups of nodes, and can be positive or negative.}
    \label{fig:hypergraph}
\end{figure}

However, a comprehensive understanding of system characterizations with high-order interactions remains elusive. High-order interactions occur among a group of individual components, and thus can have a significant aggregated effect on the system, whereas standard networks only consider pairwise connections. To address this challenge,  high-order networks (i.e., hypergraphs, see Fig.~\ref{fig:hypergraph}) and their tensorial representations are increasingly being used to model complex real-world systems, where adjacency coefficients can be defined on sets of nodes. Tensors, which are multidimensional arrays \cite{kolda2009tensor,de2000multilinear,oseledets2011tensor}, have found applications across a wide array of fields, including dynamical systems \cite{chen2019multilinear,chen2021multilinear,9903302}, signal processing \cite{sidiropoulos2017tensor,cichocki2015tensor}, network science \cite{chen2020tensor,chen2021controllability,surana2022hypergraph}, data analysis \cite{zhou2013tensor}, and more~\cite{stefansson2016sequential}. 

Not only can the static status of higher-order networks be studied, but also their long-standing interests in stability and state evolution. The associated dynamical system is thus composed of pairwise interaction matrix and high-order interaction tensors. Specifically, we are dealing with complex ecological networks, where the most commonly used model is known as the generalized Lotka--Volterra (GLV) model \cite{bhargava1989generalized,bunin2017ecological}. It is defined as
\begin{equation}
\dot{\textbf{x}} = \textbf{x} * \Big[\textbf{r} + \textbf{A}\textbf{x}\Big],
\end{equation}
where $\textbf{x}\in\mathbb{R}^n$ represents species abundance, $\textbf{r}\in\mathbb{R}^n$ is the vector of intrinsic growth rate, and $\textbf{A}\in\mathbb{R}^{n\times n}$ is the interaction matrix whose off-diagonal element $\textbf{A}_{ij}$ represents the effect of species $j$ upon species $i$. Throughout, we use $*$ to denote element-wise multiplication.

Numerous linear stability results have been developed for the GLV model by characterizing its community matrix (Jacobian matrix) \cite{may1972will, Allesina-Nature-2012}. In reality, species interactions often emerge in higher-order combinations, where the relationship between two species is influenced by one or more additional species \cite{bairey2016high}. Although the importance of high-order interactions has been recognized, their impact on the stability of ecological systems has not been fully understood. Existing findings are preliminary, most of which focus on numerical simulations \cite{grilli2017higher,singh2021higher}.

We are particularly interested in the higher-order generalized Lotka--Volterra (HOGLV) model, designed to represent higher-order interactions within intricate ecological networks. Computing the Jacobian matrices for such high-order systems is challenging due to the curse of dimensionality, i.e., the size of variables increases exponentially with the dimensionality. To address this challenge, we propose a framework, leveraging from various tensor decompositions, including higher-order singular value decomposition (HOSVD), Canonical Polyadic decomposition (CPD), and tensor train decomposition (TTD), to improve both memory and computational efficiency in the computation of Jacobian matrix for the HOGLV model.

The article is organized as follows.  In Section \ref{sec:prelim},  we begin with preliminaries on basic tensors. In Section \ref{sec:model},  we introduce the HOGLV model and represent it in the HOSVD, CPD, and TTD forms with memory complexity analyses. In Section \ref{sec:stable}, we derive the Jacobian matrix of the HOGLV model and offer a computational complexity analysis for each tensor decomposition-based representation. Finally, we conclude our work with discussions on future directions in Section \ref{sec:conclusion}.

\section{Tensor Preliminaries}\label{sec:prelim} 
The order of a tensor is defined as the number of its dimensions, with each dimension referred to as a mode. A $p$th-order tensor can be  denoted by $\mathcal{T}\in\mathbb{R}^{n_1\times n_2\times\dots\times n_p}$. The tensor vector multiplication $\mathcal{T}\times_{q}\textbf{v}$ along mode $q$ for a vector $\textbf{v}\in\mathbb{R}^{n_q}$ is defined as
$[\mathcal{T}\times_{q}\textbf{v}]_{i_1i_2\dots i_{q-1}i_{q+1}\dots i_p} := \sum_{i_q=1}^{n_q}\mathcal{T}_{i_1i_2\dots i_q\dots i_p}\textbf{v}_{i_q}.
$
The tensor matrix multiplication $\mathcal{T}\times_{q}\textbf{M}$ along mode $q$ for a matrix $\textbf{M}\in\mathbb{R}^{m\times n_q}$ is defined as
$
    [\mathcal{T}\times_{q}\textbf{M}]_{i_1i_2\dots i_{q-1}ji_{q+1}\dots i_p} := \sum_{i_q=1}^{n_q}\mathcal{T}_{i_1i_2\dots i_q\dots i_p}\textbf{M}_{ji_q}.
$ 

In the following, we introduce three key tensor decompositions utilized in this article:
\begin{itemize}
\item The higher-order singular value decomposition (HOSVD) \cite{de2000multilinear} of $\mathcal{T}\in\mathbb{R}^{n_1\times n_2\times\dots\times n_p}$ is defined as
\begin{equation}\label{eq:hosvd}
    \mathcal{T}\approx \mathcal{S}\times_1\textbf{U}_1\times_2\textbf{U}_2\times_3\cdots\times_p \textbf{U}_p,
\end{equation}
where $\mathcal{S}\in\mathbb{R}^{r_1\times r_2\times\dots\times r_p}$ represents the core tensor, and $\textbf{U}_q\in\mathbb{R}^{n_q\times r_p}$ are the factor matrices containing orthonormal columns for $q=1,2,\dots, p$. For simplicity, we may rewrite \eqref{eq:hosvd} in a more compact form, i.e.,
$
    \mathcal{T}\approx \mathcal{S}\times_{12\cdots p}\{\textbf{U}_1,\textbf{U}_2,\dots,\textbf{U}_p\}.
$
The computation of HOSVD is numerically stable and can offer a quasi-optimal approximation.

\item The Canonical Polyadic decomposition (CPD) \cite{kolda2009tensor} of $\mathcal{T}\in\mathbb{R}^{n_1\times n_2\times\dots\times n_p}$ is defined as
\begin{equation}\label{eq:cpd}
    \mathcal{T}\approx \sum_{i=1}^r [\textbf{T}_1]_{:i}\circ[\textbf{T}_2]_{:i}\circ \cdots \circ [\textbf{T}_p]_{:i},
\end{equation}
where $\textbf{T}_q\in\mathbb{R}^{n_q\times r}$ are the factor matrices for $q=1,2,\dots,p$, and $r$ is called the CP rank of $\mathcal{T}$ if it is the minimum integer achieving the decomposition. We use $\circ$ to denote the outer product. This is because CPD is not numerically stable. Moreover, the low-rank approximation of CPD is ill-posed. 

\item The tensor train decomposition (TTD) \cite{oseledets2011tensor} of $\mathcal{T}\in\mathbb{R}^{n_1\times n_2\times\dots\times n_p}$ is defined as
\begin{equation}\label{eq:ttd}
    \mathcal{T}\approx \sum_{i_0=1}^{r_0}\cdots\sum_{i_p=1}^{r_p}[\mathcal{T}_1]_{i_0:i_1}\circ [\mathcal{T}_2]_{i_1:i_2}\circ\cdots\circ[\mathcal{T}_p]_{i_{p-1}:i_p},
\end{equation}
where $\mathcal{T}_q\in\mathbb{R}^{r_{q-1}\times n_q\times r_q}$ are the third-order core tensors for $q=1,2,\dots,p$, and $\{r_0,r_1,\dots,r_p\}$ is the set of TT-ranks with $r_0=r_p=1$. Similar to HOSVD, TTD is numerically stable and its low-rank approximation is quasi-optimal. Importantly, TTD is more efficient than HOSVD in both computation and memory. 
\end{itemize}

\section{The HOGLV Model}\label{sec:model}
The dynamics of complex ecological networks with higher-order interactions are often characterized by the higher-order generalized Lotka--Volterra (HOGLV) model \cite{bairey2016high,aladwani2019addition,singh2021higher}, i.e.,
\begin{equation}\label{eq:hoglv}
\begin{split}
    \dot{x}_i &= x_i\Big [r_i + \sum_{j=1}^n \textbf{A}_{ij}x_j + \sum_{j=1}^n\sum_{k=1}^n \mathcal{B}_{ijk}x_jx_k \\
    &+ \sum_{j=1}^n\sum_{k=1}^n\sum_{l=1}^n \mathcal{C}_{ijkl}x_jx_kx_l + \cdots\Big]
\end{split}
\end{equation}
for $i=1,2,\dots, n$, where $r_i$ is the intrinsic growth rate for species $i$, $\textbf{A}\in\mathbb{R}^{n\times n}$ is the interaction matrix whose off-diagonal elements $\textbf{A}_{ij}$ represent the effect of species $j$  upon species $i$, $\mathcal{B}\in\mathbb{R}^{n\times n\times n}$ is the third-order interaction tensor whose off-diagonal elements represent the effect that species $j$ and $k$ has upon species $i$, and $\mathcal{C}\in\mathbb{R}^{n\times n\times n\times n}$ is the fourth-order interaction tensor whose off-diagonal elements represent the effect that species $j$, $k$, $l$ has upon species $i$. In fact, the HOGLV model belongs to the family of polynomial dynamical systems. 
Alternatively, we may rewrite the system of the differential equations (\ref{eq:hoglv}) into the vector form using tensor products, which is 
$
     \dot{\textbf{x}} = \textbf{x} * [\textbf{r} + \textbf{A}\textbf{x} + \mathcal{B}\times_{23}\{\textbf{x},\textbf{x}\} + \mathcal{C}\times_{234}\{\textbf{x},\textbf{x},\textbf{x}\}+ \cdots ].
$
The total number of the model parameters in HOGLV model, whose maximum interaction order is $M$, can be estimated as $\text{MC}_{\text{full}} \sim \mathcal{O} (n^M)$. It can be directly seen that the $M$th-order interaction tensor has at most $n^M$ entries. Obviously, the memory complexity scales exponentially with the maximum order of interactions $M$. For large ecological networks with higher-order interactions, it will be challenging to handle the full representation due to severe memory limitations. In the following, we represent the HOGLV model in various tensor decomposition forms.

\subsection{HOSVD-based Representation}
We first consider applying HOSVD  to the interaction tensors to derive a HOSVD-based representation of the HOGLV model.

\textit{Proposition 1:} Suppose that HOSVDs of interaction tensors $\mathcal{B}\in\mathbb{R}^{n\times n\times n}$ and $\mathcal{C}\in\mathbb{R}^{n\times n\times n\times n}$ are provided as
$
        \mathcal{B} \approx \mathcal{B}_0\times_{123} \{\textbf{B}_1,\textbf{B}_2,\textbf{B}_3\} \text{ and }
        \mathcal{C} \approx \mathcal{C}_0\times_{1234} \{\textbf{C}_1,\textbf{C}_2,\textbf{C}_3,\textbf{C}_4\},
$
where $\mathcal{B}_0\in\mathbb{R}^{b_1\times b_2\times b_3}$ and $\mathcal{C}_0\in\mathbb{R}^{c_1\times c_2\times c_3\times c_4}$ are the core tensors with associated factor matrices $\textbf{B}_i\in\mathbb{R}^{n\times b_i}$ and $\textbf{C}_i\in\mathbb{R}^{n\times c_i}$, respectively. The HOSVD-based  representation of the HOGLV model can be computed as
\begin{equation}\label{eq:hosvd-hoglv}
\begin{split}
     \dot{\textbf{x}} &= \textbf{x} * \Big[\textbf{r} + \textbf{A}\textbf{x} + \mathcal{B}_0\times_{123}\{\textbf{B}_1,\textbf{B}_2^\top\textbf{x},\textbf{B}_3^\top\textbf{x}\} \\ & + \mathcal{C}_0\times_{1234}\{\textbf{C}_1,\textbf{C}_2^\top\textbf{x},\textbf{C}_3^\top\textbf{x},\textbf{C}_4^\top\textbf{x}\}+ \cdots  \Big ].
\end{split}
\end{equation}

\begin{proof}
According to the properties of tensor vector/matrix products and HOSVD, it can be shown that
\begin{equation*}
\begin{split}
    &\Big(\mathcal{B}_0\times_{123} \{\textbf{B}_1,\textbf{B}_2,\textbf{B}_3\}\Big)\times_{23}\{\textbf{x},\textbf{x}\}\\
    & = \mathcal{B}_0\times_{123}\{\textbf{B}_1,\textbf{B}_2^\top\textbf{x},\textbf{B}_3^\top\textbf{x}\}.
\end{split}
\end{equation*}
Similarly, the same principle applies to the interaction tensor $\mathcal{C}$ as well. Thus, the result follows immediately.    
\end{proof}

\textit{Remark 1:} Suppose that the reduced dimensions of the interaction tensors are all equal to $r$, i.e., $b_i=c_i=\cdots=r$. If the maximum order of interactions is $M$, the total number of parameters in the HOSVD-based representation of the HOGLV model can be estimated as
\begin{equation*}
\text{MC}_{\text{hosvd}}\sim \mathcal{O}(n^2+ M^2nr + r^M).
\end{equation*}

While the above memory complexity increases exponentially with the maximum order of interactions $M$, it is mitigated by the fact that the reduced dimension $r$ can be small for certain structured tensors.

\subsection{CPD-based Representation}
We explore applying CPD to the interaction tensors to obtain a CPD-based representation of the HOGLV model.

\textit{Proposition 2:} Suppose that the CPDs of the interaction tensors $\mathcal{B}\in\mathbb{R}^{n\times n\times n}$ and $\mathcal{C}\in\mathbb{R}^{n\times n\times n\times n}$ are provided as 
$
        \mathcal{B} \approx \sum_{i=1}^{b}[\textbf{B}_1]_{:i}\circ [\textbf{B}_2]_{:i}\circ [\textbf{B}_3]_{:i} \text{ and }
        \mathcal{C} \approx \sum_{i=1}^{c}[\textbf{C}_1]_{:i}\circ [\textbf{C}_2]_{:i}\circ [\textbf{C}_3]_{:i}\circ [\textbf{C}_4]_{:i},
$
where $\textbf{B}_i\in\mathbb{R}^{n\times b}$ and $\textbf{C}_i\in\mathbb{R}^{n\times c}$ are the factor matrices. The CPD-based representation of the HOGLV model can be computed as
\begin{equation}\label{eq:cpd-hoglv}
\begin{split}
     \dot{\textbf{x}} &= \textbf{x} * \Big[\textbf{r} + \textbf{A}\textbf{x} + \sum_{i=1}^b\big([\textbf{B}_2]_{:i}^\top\textbf{x}\big)\big([\textbf{B}_3]_{:i}^\top\textbf{x}\big)[\textbf{B}_1]_{:i}\\
     +& \sum_{i=1}^c \big([\textbf{C}_2]_{:i}^\top\textbf{x}\big)\big([\textbf{C}_3]_{:i}^\top\textbf{x}\big)
     \big([\textbf{C}_4]_{:i}^\top\textbf{x}\big)[\textbf{C}_1]_{:i} +\cdots\Big].
\end{split}
\end{equation}

\begin{proof}
From the properties of tensor vector/matrix products and CPD, it can be shown that
\begin{equation*}
\begin{split}
    & \Big(\sum_{i=1}^b [\textbf{B}_1]_{:i}\circ [\textbf{B}_2]_{:i}\circ [\textbf{B}_3]_{:i}\Big)\times_{23}\{\textbf{x}, \textbf{x}\}\\
    & = \sum_{i=1}^b\big([\textbf{B}_2]_{:i}^\top\textbf{x}\big)\big([\textbf{B}_3]_{:i}^\top\textbf{x}\big)[\textbf{B}_1]_{:i}.
\end{split}
\end{equation*}
Likewise, the same principle applies to the interaction tensor $\mathcal{C}$ as well. Therefore, the result follows immediately.    
\end{proof}

\textit{Remark 2:} Suppose that the CP ranks of the interaction tensors are all equal to $r$, i.e., $b=c=\dots=r$. If the maximum order of interactions is $M$, the total number of parameters in the CPD-based representation of the HOGLV model can be estimated as 
\begin{equation*}
    \text{MC}_{\text{cpd}}\sim \mathcal{O}(n^2+M^2nr).
\end{equation*}

Remarkably, the above memory complexity increases quadratically with both the maximum order of interactions $M$ and the number of species $n$. Furthermore, when the CP rank $r$ is small, the CPD-based representation of the HOGLV model can significantly enhance memory efficiency. However, it is important to note that CPD may not be numerically stable, which may limit the applicability of this representation.

\subsection{TTD-based Representation}
We investigate the use of TTD on the interaction tensors to derive a TTD-based representation of the HOGLV model.

\textit{Proposition 3:} Suppose that the TTDs of the interaction tensors $\mathcal{B}\in\mathbb{R}^{n\times n\times n}$ and $\mathcal{C}\in\mathbb{R}^{n\times n\times n\times n}$ are provided as
$
        \mathcal{B} \approx \sum_{i_0=1}^{b_0}\cdots \sum_{i_3=1}^{b_3}[\mathcal{B}_1]_{i_0:i_1}\circ [\mathcal{B}_2]_{i_1:i_2}\circ [\mathcal{B}_3]_{i_2:i_3} \text{ and }
        \mathcal{C} \approx \sum_{i_0=1}^{c_0}\cdots \sum_{i_4=1}^{c_4}[\mathcal{C}_1]_{i_0:i_1}\circ [\mathcal{C}_2]_{i_1:i_2}\circ [\mathcal{C}_3]_{i_2:i_3}\circ [\mathcal{C}_4]_{i_3:i_4},
$
where $\mathcal{B}_i\in\mathbb{R}^{b_{i-1}\times n\times b_i}$ and $\mathcal{C}_i\in\mathbb{R}^{c_{i-1}\times n\times c_i}$ are the third-order core tensors. The TTD-based representation of the HOGLV model can be computed as
\begin{equation}\label{eq:ttd-hoglv}
\small
\begin{split}
     \dot{\textbf{x}} &= \textbf{x} * \Big[\textbf{r} + \textbf{A}\textbf{x} + \sum_{i_0=1}^{b_0}\cdots \sum_{i_3=1}^{b_3}([\mathcal{B}_2]_{i_1:i_2}^\top\textbf{x})([\mathcal{B}_3]_{i_2:i_3}^\top\textbf{x})[\mathcal{B}_1]_{i_0:i_1}\\
     + & \sum_{i_0=1}^{c_0}\cdots \sum_{i_4=1}^{c_4}([\mathcal{C}_2]_{i_1:i_2}^\top\textbf{x})([\mathcal{C}_3]_{i_2:i_3}^\top\textbf{x})
     ([\mathcal{C}_4]_{i_3:i_4}^\top\textbf{x})[\mathcal{C}_1]_{i_0:i_1}+\cdots\Big].
\end{split}
\end{equation}
\begin{proof}
The proof is similar to Proposition 2.    
\end{proof}

\textit{Remark 3:} Suppose that the TT-ranks of the interaction tensors are all equal to $r$, i.e., $b_i=c_i=\cdots=r$. Note that the first and last TT-ranks of each tensor are always equal to 1. If the maximum order of interactions is $M$, the total number of parameters in the TTD-based representation of the HOGLV model can be estimated as
\begin{equation*}
    \text{MC}_{\text{ttd}}\sim \mathcal{O}(n^2+ M^2nr^2).
\end{equation*}

While the above memory complexity is slightly higher than that of the CPD-based representation, it remains significantly more efficient than the full and HOSVD-based representations. Importantly, TTD can offer the crucial advantages of numerical stability and quasi-optimal approximation.

\section{Complexity Analyses of Linear Stability}\label{sec:stable}
Linear stability analysis has proven to be a successful tool in the study of complex ecological networks, enhancing our understanding of the intricate relationships between stability and biodiversity \cite{allesina2012stability,allesina2015stability,may1972will}. A plethora of research has focused on the linear stability of the GLV model \cite{gibbs2018effect,baron2023breakdown}. The essence of linear stability lies in the computation of the Jacobian matrix. Here, we derive the Jacobian matrix for the HOGLV model and perform a computational complexity analysis of computing the Jacobian matrix using the full, HOSVD-based, CPD-based, and TTD-based representations.

\textit{Proposition 4:} Suppose that the equilibrium point of the HOGLV model is $\textbf{x}^{*}$. The Jacobian matrix of the HOGLV model evaluated at $\textbf{x}^{*}$ can be computed as
\begin{equation}\label{eq:jacobian}
\begin{split}
    \textbf{M} &= \textbf{X}^{*}\Big[\textbf{A} + \mathcal{B}\times_2\textbf{x}^{*} + \mathcal{B}\times_3\textbf{x}^{*} + \mathcal{C}\times_{23}\{\textbf{x}^{*},\textbf{x}^{*}\}\\
    &+\mathcal{C}\times_{24}\{\textbf{x}^{*},\textbf{x}^{*}\} + \mathcal{C}\times_{34}\{\textbf{x}^{*},\textbf{x}^{*}\}+\cdots \Big],
\end{split}
\end{equation}
where $\textbf{X}^{*}\in\mathbb{R}^{n\times n}$ is a diagonal matrix that contains $\textbf{x}^{*}$ along its diagonal.

\begin{proof}
Based on the properties of tensor vector products, it can be demonstrated that 
\begin{equation*}
    \frac{d}{d\textbf{x}}\mathcal{B}\times_{23}\{\textbf{x},\textbf{x}\} = \mathcal{B}\times_2\textbf{x} + \mathcal{B}\times_3\textbf{x}. 
\end{equation*}
Likewise, the same derivative principle is applicable to the interaction tensor $\mathcal{C}$. Consequently, the result can be derived in a manner similar to the computation of the Jacobian matrix for the GLV model~\cite{bunin2017ecological}.  
\end{proof}

\textit{Remark 4:} If the maximum order of interactions is $M$, the computational complexity of computing the Jacobian matrix of the HOGLV model can be estimated as
\begin{equation*}
    \text{CC}_{\text{full}} \sim \mathcal{O}(M^2n^M).
\end{equation*}

Similar to its memory complexity, the computational complexity of the full representation scales exponentially with the maximum order of interactions $M$. Therefore, computing the Jacobian matrix for large ecological networks with higher-order interactions is an exceptionally challenging task. To address this challenge, we can utilize tensor decomposition-based representations of the HOGLV model, which significantly reduces the computational complexity. In the following, we estimate the computational complexity of computing the Jacobian matrix using the HOSVD-based, CPD-based, and TTD-based representations. 

\textit{Remark 5:} Given the HOSVD-based representation of the HOGLV model (\ref{eq:hosvd-hoglv}) with reduced dimension $r$ for all interaction  tensors, if the maximum order of interactions is $M$, the computational complexity of computing the Jacobian matrix  can be estimated as
\begin{equation*}
    \text{CC}_{\text{hosvd}} \sim \mathcal{O}(M^2n^2r+M^3nr+M^2nr^M).
\end{equation*}

\textit{Remark 6:} Given the CPD-based representation of the HOGLV model (\ref{eq:cpd-hoglv}) with CP rank $r$ for all interaction tensors, if the maximum order of interactions is $M$, the computational complexity of computing the Jacobian matrix  can be estimated as
\begin{equation*}
    \text{CC}_{\text{cpd}} \sim \mathcal{O}(M^2n^2r+M^3nr).
\end{equation*}

\textit{Remark 7:} Given the TTD-based representation of the HOGLV model (\ref{eq:ttd-hoglv}) with TT-rank $r$ for all interaction tensors (except for the first and last TT-ranks), if the maximum order of interactions is $M$, the computational complexity of computing the Jacobian matrix  can be estimated as
\begin{equation*}
    \text{CC}_{\text{ttd}} \sim \mathcal{O}(Mnr^M+n^2r^M).
\end{equation*}

The proofs of the above remarks can be found in the appendix. The computational complexity of computing the Jacobian matrix using the HOSVD-based, CPD-based, and TTD-based representations exhibits a similar behavior to their memory complexity. Particularly, the CPD-based representation stands out with the lowest computational complexity, which does not grow exponentially with the maximum order of interactions $M$. Nonetheless, accurately computing CPD becomes challenging for tensors with larger dimensions. Conversely, the TTD-based representation, although having a higher computational complexity compared to the CPD-based representation, outperforms the full and HOSVD-based representations with guaranteed numerical stability.

After obtaining the Jacobian matrix, we can compute its eigenvalues to determine the linear stability of the HOGLV model at an equilibrium point. It is important to acknowledge that computing the eigenvalues of the Jacobian matrix can be computationally demanding for large ecological networks, but this specific aspect is not the primary focus of this article. Nevertheless, recent research \cite{mach2012computing, dolgov2014computation} has explored the use of TTD to expedite eigenvalue computations, which can be useful here for future investigations.

\section{Numerical Examples}\label{sec:num}
We present three case studies to illustrate our framework. All three examples were conducted on a Macintosh machine equipped with 32 GB RAM and an Apple M1 Pro chip  (10-core CPU at 3.2 GHz, 16-core GPU, and 16-core Neural Engine) using MATLAB R2022a with the  Tensor Toolbox 3.0 \cite{kola2017tensor} and TT Toolbox \cite{oseledets2016tt}.

\begin{figure}[t]
    \centering
    \includegraphics[width=0.43\textwidth]{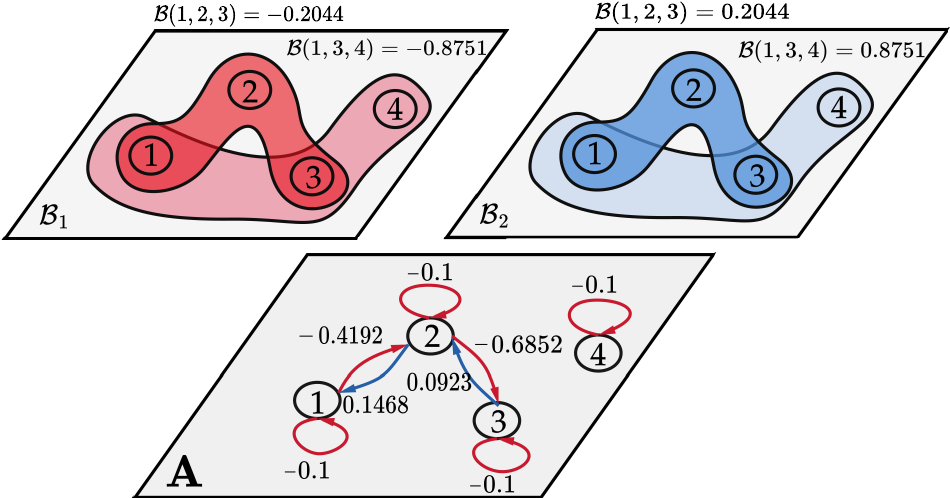}
    \caption{Two ecological networks, fully characterized by their pairwise and third-order interactions, with positive interactions colored in blue and negative interactions in red.}
    \label{fig:unstable}
\end{figure}
\subsection{Computing Jacobian Matrices}
We computed the Jacobian matrices of two ecological networks captured by the HOGLV model using \eqref{eq:jacobian}. For simplicity, we set the total number of species and the maximum order of interactions to 4 and  3, respectively. The structures of the two ecological networks are presented in Fig. \ref{fig:unstable}, which share pairwise interactions. 
Suppose that the equilibrium point is $\textbf{x}^{*}=\begin{bmatrix}
   1 & 1 & 1 & 1
\end{bmatrix}^\top$ for both networks. The associated Jacobian matrices can be computed as 
\begin{align*}
\textbf{M}_1&=\begin{bmatrix}
    -0.1000 &  -0.2147 &\phantom{-}1.0826  &\phantom{-}0.8781\\
    \phantom{-}0.1468 &  -0.1000 &  -0.6852    &\phantom{-}0\\
    \phantom{-}0&  \phantom{-}0.0923  & -0.1000    &\phantom{-}0\\
    \phantom{-}0  & \phantom{-}0   &\phantom{-} 0  & -0.1000
    \end{bmatrix},\\
    \textbf{M}_2&=\begin{bmatrix}
    -0.1000 &  -0.6236 &  -1.0826  & -0.8781\\
    \phantom{-}0.1468  & -0.1000 &  -0.6852    &\phantom{-}0\\
    \phantom{-}    0  &  \phantom{-}0.0923  & -0.1000  &\phantom{-}0\\
    \phantom{-}    0   & \phantom{-}0   & \phantom{-}0  & -0.1000
    \end{bmatrix}.
\end{align*}
The first Jacobian matrix $\textbf{M}_1$ has one positive eigenvalue, indicating that the first ecological network is locally unstable around the equilibrium point. The second Jacobian matrix $\textbf{M}_2$ has all eigenvalues negative, suggesting that the second ecological network is locally stable near the equilibrium point. This example demonstrates that higher-order interactions can significantly influence the stability of complex ecological networks.

\subsection{Memory Comparison}
We conducted a comparison of the total number of model parameters for the full, HOSVD-based, CPD-based, and TTD-based representations of the HOGLV model using random sparse tensors with varying dimensions. We set the maximum order of interactions to 4 and randomly generated sparse interaction tensors $\mathcal{B}\in\mathbb{R}^{n\times n\times n}$ and $\mathcal{C}\in\mathbb{R}^{n\times n\times n\times n}$ for $n=10,20,\dots,50$. By employing the HOSVD, CPD, and TTD on the interaction tensors, we computed their corresponding representations of the HOGLV model. The results are presented in Table \ref{tab:1}. The HOSVD-based representation is effective in reducing the number of model parameters for relatively small dimensions ($n=10,20$). However, as the dimension increases, its number of model parameters surpasses that of the full representation. Moreover, the CPD-based representation demonstrates the lowest memory usage. However, we failed to accurately compute the CPDs of the interaction tensors for larger dimensions ($n=30,40,50$) due to the issue of numerical instability. Thus, the TTD-based representation can be considered as a robust and viable choice, which maintains a reasonably low number of model parameters compared to the full representation.

\begin{table}[t]
\centering
\caption{Comparison of the total number of model parameters for the full, HOSVD-based, CPD-based, and TTD-based representations of the HOGLV model across varying system dimensions.}
\begin{tabular}{|l|l|l|l|l|l|}
\hline
Dimension            & $10$    & $20$    & $30$    & $40$    & $50$    \\ \hline
Full        & 11,110 & 168,420 & 837,930 & 2,625,640 & 6,377,550 \\ \hline
HOSVD & 810  & 131,608 & 844,230 & 2,636,840 & 6,395,050 \\ \hline
CPD   & 460  & 6,860  & -  & - & -  \\ \hline
TTD   & 890  & 36,880  & 238,530 & 920,040 & 2,687,550  \\ \hline
\end{tabular}
\label{tab:1}
\end{table}

\subsection{Computation Comparison}
\begin{figure}[t]
    \centering
    \includegraphics[scale=0.38]{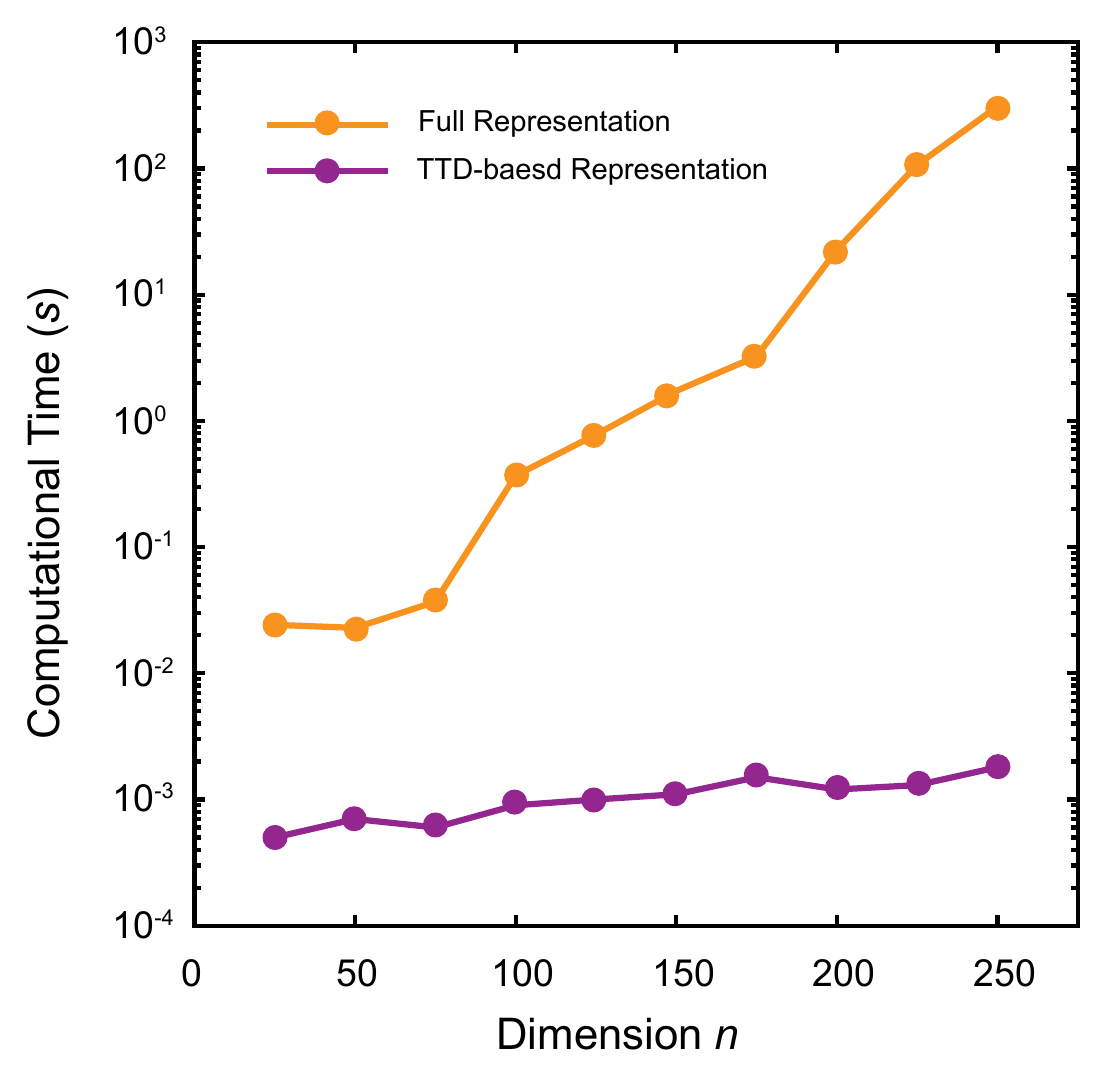}
    \caption{Comparison of computational time for computing the Jacobian matrix between the full and TTD-based representations.}
    \label{fig:1}
\end{figure}

We compared the computational time required for computing the Jacobian matrix of the HOGLV model using the full representation and the TTD-based representation. We exclusively considered the TTD-based representation here due to its numerical stability and low computational complexity. More importantly, it benefits from a well-established TT-algebra, making it a suitable choice for this computation. We set the maximum order of interactions to 4 and randomly generated interaction tensors $\mathcal{B}\in\mathbb{R}^{n\times n\times n}$ and $\mathcal{C}\in\mathbb{R}^{n\times n\times n\times n}$ in the TTD format with low TT-ranks for varying dimensions. In other words, we assumed that the HOGVL model is initially provided in the TTD-based representation. The findings are displayed in Fig. \ref{fig:1}. Evidently, the TTD-based representation offers a substantial time advantage in computing the Jacobian matrix compared to the full representation. It is noteworthy that the full representation encounters memory limitations for $n\geq 275$, which fails to complete the computation. On the other hand, the TTD-based representation maintains computational efficiency even for larger dimensions, as demonstrated in Table \ref{tab:2}.

\begin{table}[t]
\centering
\caption{Computation time for calculating the Jacobian matrix of the TTD-based representation of the HOGLV model for large system dimensions, where the full representation fails due to memory constraints.}
\begin{tabular}{|l|l|l|l|l|l|}
\hline
Dimension  & $1000$ & $2000$ & $3000$ & $4000$ & $5000$ \\\hline
Time (s)     &  0.0259     &  0.1675    &   0.5304   &  1.1228    &  1.9990 \\\hline    
\end{tabular}
\label{tab:2}
\end{table}

\section{Conclusion}\label{sec:conclusion}
In this article, we proposed a framework to tackle the problem of complex ecological systems with high-order interactions. In particular, we were interested in determining the linear stability of the HOGLV model by computing the Jacobian matrix and its eigenvalues. A crucial challenge in the analysis of complex systems and coupled representations is their exponentially increasing size.  In our approach, we explored various tensor decomposition techniques for good scalability with detailed memory and computational complexity investigations. 

The stability analysis of dynamical systems with high-order interactions is not limited to ecological systems, but also appears in various fields, such as opinion dynamics and social networks, where the aggregated impact of high-order interactions plays a significant role.  
Another area of interest is the estimation of interaction matrices and tensors of different orders \cite{dong2023data}. Additionally, the study of high-order time-varying systems, where the interaction tenors are time-varying, is a promising direction for future research.

\bibliographystyle{plain}
\bibliography{references}

\appendix 
We provide detailed proofs for Remarks 4 and 5 here. Remarks 6 and 7 can be proved in a similar manner. 

\section*{Proof of Remark 4}
We first consider the computational complexity of tensor vector multiplications, which can be estimated as
\begin{equation*}
    \mathcal{O}(2n^3+3n^4+\cdots \binom{M}{2}n^M) \sim \mathcal{O}(M^2n^M).
\end{equation*}
Additionally,  the computational complexity of matrix additions and multiplications is much less than $\mathcal{O}(M^2n^M)$. Therefore, the overall complexity of computing the Jacobian matrix of the HOGLV model using the full representation is about $\mathcal{O}(M^2n^M)$.

\section*{Proof of Remark 5}
Suppose that the reduced dimensions of the interaction tensors are all equal to $r$, i.e., $b_i=c_i=\cdots=r$. We first estimate the computational complexity of computing  the term $\mathcal{B}_0\times_{123}\{\textbf{B}_1,\textbf{B}_2^\top\textbf{x},\textbf{B}_3\}$, which is about 
\begin{equation*}
    \mathcal{O}(nr+nr^3+nr^2+n^2r).
\end{equation*}
Therefore, the computational complexity of tensor vector multiplications can be approximated by
\begin{align*}
    &2\mathcal{O}(nr+nr^3+nr^2+n^2r) \\
    +&3\mathcal{O}(2nr+nr^2+nr^3+nr^4+n^2r)\\ 
    +&\cdots\\ 
    +&\binom{M}{2}\mathcal{O}((M-1)nr+nr^2+\cdots + nr^M+n^2r)\\
    &\sim\mathcal{O}(M^3nr+M^2nr^M+M^2n^2r).
\end{align*}
Again, the computational complexity of matrix additions and multiplications can be negligible. Thus, the overall complexity of computing the Jacobian matrix of the HOGLV model using the HOSVD representation is about $\mathcal{O}(M^3nr+M^2nr^M+M^2n^2r)$. 
\end{document}